\documentclass[fleqn,10pt]{wlscirep}

\usepackage{color}
\definecolor{MYyellow}{RGB}{255,190,0}
\definecolor{MYgreen}{RGB}{0,200,120}

\title{Faithful quantum state transfer between weakly coupled qubits}

\author[1]{M. Mikov\'{a}}

\author[1]{I. Straka}

\author[1]{M. Mi\v{c}uda}

\author[1]{V. Kr\v{c}marsk\'{y}}

\author[1]{M. Du\v{s}ek}

\author[1]{M. Je\v{z}ek}

\author[1]{J. Fiur\' a\v sek}

\author[1,*]{R. Filip}

\affil[1]{Department of Optics, Faculty of Science, Palack\' y University, 17. listopadu 1192/12,  771~46 Olomouc,  Czech Republic}

\affil[*]{filip@optics.upol.cz}

\begin{abstract}
One of the strengths of quantum information theory is that it can treat quantum states without referring to their particular physical representation. 
In principle, quantum states can be therefore fully swapped between various quantum systems by their mutual interaction and this quantum state transfer is crucial  
for many quantum communication and information processing tasks. In practice, however, the achievable interaction time and strength are often limited by decoherence. 
Here we propose and experimentally demonstrate a procedure for faithful quantum state transfer between two weakly interacting qubits. Our scheme enables a probabilistic 
yet perfect unidirectional transfer of an arbitrary unknown state of a source qubit onto a target qubit prepared initially in a known state. The transfer is achieved by 
a combination of a suitable measurement of the source qubit and quantum filtering on the target qubit depending on the outcome of measurement on the source qubit. 
We experimentally verify feasibility and robustness of the transfer using a linear optical setup with qubits encoded into polarization states of single photons.
\end{abstract}
\begin{document}

\flushbottom
\maketitle

\thispagestyle{empty}

\section*{Introduction}

A full exploitation of the potential of quantum information theory \cite{Nielsen00,Macchiavello00} requires development of hybrid quantum information processing devices optimally 
combining and interconnecting various physical platforms. For instance, photonic qubits are ideal carries of quantum information in quantum communication \cite{Gisin07},
while atomic, solid state or superconducting stationary qubits are well suited for local storage and processing of quantum information \cite{Haffner08,Kimble08,Saffman10,Plantenberg07}.
The future of quantum technologies therefore crucially depends on our ability to efficiently  interconnect quantum states between different 
physical platforms \cite{Kuzmich03,Wal03,Wilk07,Yilmaz10,Stute13,Bernien13,Pirkkalainen13,Tiecke14,Tan15,Ballance15,Riedinger16}.
However, the qubits often interact only weakly, or their strong interaction is affected by decoherence \cite{Zurek03}. The decoherence therefore forces us to limit the interaction time
which typically leads to weak coupling.
A weak coupling prevents us to directly implement a full bidirectional swap of quantum states of the two qubits \cite{You00}, and part of the information remains not transferred 
after the interaction.

For the one-way inter-connection of two physical platforms, a high quality unidirectional state transfer from a source qubit to a target qubit could be sufficient.
Here we propose and experimentally verify a method for the unidirectional qubit-state transfer via an arbitrary weak purity-preserving coupling between the qubits. 
The scheme is universally applicable with potential to connect various platforms and systems \cite{Wallquist09,Xiang13}.
The protocol combines a suitable projective measurement on the source qubit with optimal quantum filter \cite{Verstraete01} on the target, 
and a real-time feed-forward \cite{Giacomini02,Prevedel07,Mikova12,Mikova13,Pfaff14,Steffen13,Riebe04}. 
Advantageously, the initial state of the target qubit system could be its ground state that is minimally affected by the environment. 
Our approach where weak qubit-qubit interaction is combined with local operations on each qubit and exchange of
classical information enables perfect state transfer, although generally only in a probabilistic manner. Such perfect heralded conditional 
state transfer may be useful and sufficient in a variety of situations. For instance, quantum entanglement can be conditionally transferred 
and then teleportation \cite{Riebe04,Bennett93,Bouwmeester97,Barrett04} can provide a deterministic transfer later on.

\begin{figure}[t]
\centerline{\includegraphics[width=0.6\textwidth]{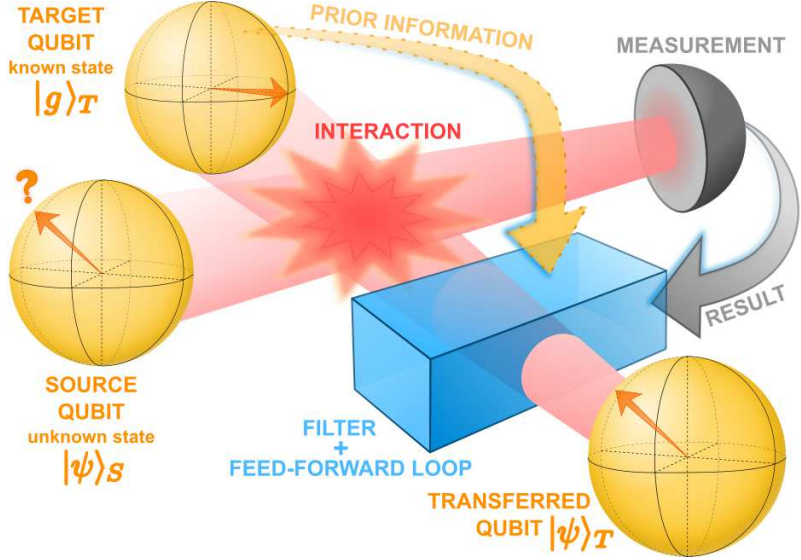}}
\vspace*{3mm}
	\caption{Quantum state transfer protocol. The initial state $|\psi\rangle_S$ of the source qubit $S$ can be arbitrary and unknown. 
	The goal is to transfer any state of source $S$ to target $T$, initially in a known fixed state $|g\rangle_T$. 
		The state transfer requires some interaction (unitary or probabilistic) between the source $S$ and target $T$ qubits.  
		The transfer is performed by optimal measurement and feed-forward loop with optimal filter controlled by the measurement result 
		and prior information about the initial target state $|g\rangle_T$.}
	\label{fig:protocol}
\end{figure}

\section*{Results}

\subsection*{Single qubit transfer} 
The goal of the universal quantum state transfer protocol is to faithfully map any quantum state $|\psi\rangle_S=\alpha |0\rangle_S+\beta|1\rangle_S$ 
of a source qubit $S$ onto the target qubit $T$ that is initially prepared in a known fixed state $|g\rangle_T$, as illustrated in Fig.~\ref{fig:protocol}.  
The source qubit could even be initially entangled with some ancilla qubit.  For clarity of subsequent presentation, we shall consider 
a generic pure initial state of the source qubit  $|\psi\rangle_S$. If the state transfer protocol works perfectly for all pure input states then, by linearity,
it would work also for mixed states or parts of entangled states. To make our treatment sufficiently general, we allow for both deterministic
and probabilistic interactions $\hat{V}$ between source and target qubits, hence $\hat{V}$ can be either a unitary operation or a non-unitary quantum filter
satisfying $\hat{V}^\dagger \hat{V}\leq \hat{I}$. We thus consider the most general class of noiseless quantum interactions.
The interaction $\hat{V}$ creates an entangled state of source and target qubits,
\begin{equation}
\hat{V}|\psi\rangle_S|g\rangle_T = \alpha|\Phi_{0}\rangle_{ST}+\beta |\Phi_{1}\rangle_{ST},
\label{PhiST}
\end{equation}
where  $|\Phi_{j}\rangle_{ST}=\hat{V}|j\rangle_S|g\rangle_T$. In the next step of the protocol we erase the correlations between soruce and target qubits
 by a projective measurement on the source qubit.
If we project the source qubit onto a pure state $|\pi\rangle_S$, we prepare the target qubit in the following pure state,
\begin{equation}
|\varphi\rangle_T=\alpha |\phi_0\rangle_T+\beta|\phi_1\rangle_T,
\label{phiT}
\end{equation}
where $|\phi_{j}\rangle_T=_S\!\!\langle \pi|\Phi_j\rangle_{ST}$. Note that the states $|\phi_0\rangle$ and $|\phi_{1}\rangle$ 
are generally non-orthogonal, $\langle\phi_0|\phi_1\rangle \neq 0$, 
and they are not normalized and their norms can differ, $\langle \phi_0|\phi_0\rangle \neq \langle \phi_1|\phi_1\rangle$.

To complete the quantum state transfer we need to transform the two non-orthogonal states $|\phi_0\rangle$ and  $|\phi_1\rangle$ 
onto normalized orthogonal basis states $|0\rangle$ and $|1\rangle$, respectively. Provided that $|\phi_0\rangle$ and $|\phi_1\rangle$ are linearly independent, 
this can be accomplished by a suitable quantum filter $\hat{G}$ applied to the target qubit,
\begin{equation}
\hat{G}=\frac{1}{N}\left(\frac{1}{\langle\phi_1^{\perp}|\phi_0\rangle}|0\rangle\langle\phi_1^{\perp}|+\frac{1}{\langle\phi_0^{\perp}|\phi_1\rangle}|1\rangle\langle\phi_0^{\perp}|\right).
\end{equation}
Here $|\phi_j^\perp\rangle$ denotes a qubit state orthogonal to $|\phi_j\rangle$, $\langle \phi_j^\perp|\phi_j\rangle=0$, and $N$ is a normalization factor. For more details, see Methods. To reach maximal probability of success, $N$ has to be set such that
the maximum singular value of $\hat{G}$ is equal to $1$ and $\hat{G}^\dagger \hat{G} \leq \hat{I}$ is satisfied.  We emphasize that $\hat{G}$ does not depend
on the input state $|\psi\rangle_S$ of the source qubit, it depends only on the initial state of the target qubit $|g\rangle_T$,
the interaction $\hat{V}$, and the state $|\pi\rangle_S$ onto which the source qubit is projected. After filtering, the state of the target qubit becomes equal to the input state of the source,
\begin{equation}
\hat{G} |\varphi\rangle_T=\frac{1}{N}(\alpha|0\rangle+\beta|1\rangle)_T,
\end{equation}
and the probability $p$ of success of the transfer protocol reads $1/|N|^2$. The probability of success can be 
maximized by optimization of the measurement strategy and enhanced by a feed-forward loop, 
which allows us to exploit both outcomes of projective measurement on the source qubit (see Methods).

 \subsection*{Example} To illustrate our method, we consider as an instructive example a class of symmetric probabilistic two-qubit interactions described by an operator 
 $\hat{V}$ diagonal in the computational basis,
 \begin{equation}
 \hat{V}=|00\rangle\langle 00|+ t_1|01\rangle \langle 01|+t_1|10\rangle\langle 10|+t_{11}|11\rangle \langle 11|,
 \label{VPPBS}
 \end{equation}
 where $t_1,t_{11}\in\langle -1,1\rangle$. This conditional interaction represents a generalized imperfect version of 
 a two-qubit quantum parity check \cite{Pittman02,Hofmann02,Okamoto09} with a weak interaction strength. 
 It turns out that in this case it is advantageous to measure the source qubit in the balanced superposition basis $|\pm\rangle=\frac{1}{\sqrt{2}}(|0\rangle \pm|1\rangle)$.
  After some algebra we find that the conditional states of the target qubit corresponding to these two outcomes differ only by a sign in the superposition,
 \begin{equation}
 |\varphi_{\pm}\rangle_T= \frac{1}{\sqrt{2}}(\alpha|\phi_0\rangle \pm \beta |\phi_1\rangle)_T,
 \label{phiTpm}
 \end{equation}
 where $|\phi_j\rangle_T= \hat{W}_j|g\rangle_T$  and
 \begin{equation}
 \hat{W}_{0}= |0\rangle\langle 0|+t_1 |1\rangle \langle 1|, \quad
  \hat{W}_{1}= t_1|0\rangle\langle 0|+t_{11}|1\rangle \langle 1|.
 \end{equation}
 Formula (\ref{phiTpm}) implies that the quantum filters $\hat{G}_{+}$ and $\hat{G}_{-}$ associated with the measurement outcomes $+$ and $-$ differ only by a fixed unitary transformation,
 $\hat{G}_{-}=\hat{U}_{\pi}\hat{G}_{+}$, where $\hat{U}_{\pi}= |0\rangle\langle 0|-|1\rangle \langle 1|$. The whole protocol can be therefore implemented using a \emph{fixed}
 quantum filter $\hat{G}_{+}$ followed by a feed-forward-controlled unitary phase shift. This greatly simplifies the experimental implementation of the protocol.

\begin{figure}[!t]
 	\centerline{\includegraphics[width=0.6\textwidth]{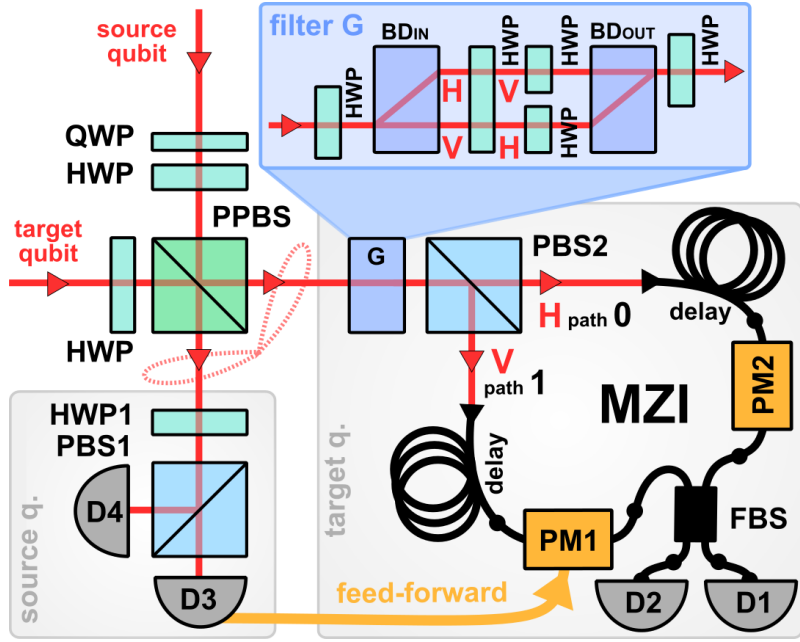}}
 	\vspace*{3mm}
 	\caption{Schematic of the experimental setup. Polarization states of photons are controlled with the help of half-wave plates (HWP) and a quarter-wave plate (QWP).
 		The photons interfere on a partially polarizing bulk beam splitter PPBS and the polarization state of source photon is measured using a HWP1, polarizing beam splitter PBS1,
 		and two single photon detectors D3 and D4. A tunable polarization filter G is constructed from two calcite beam displacers BD, and wave plates.
 		After filtration, the target photon is coupled into fiber-based Mach-Zehnder interferometer formed by two polarization maintaining
 		fibers and a balanced fiber coupler FBS. Real-time feed-forward is implemented by
 		driving one of the  integrated  phase modulators (PM1) by an electronic signal from detector D3.
 	}
 	\label{fig:schema}
 \end{figure}

 \subsection*{Experimental test} 
It is important to verify the feasibility, robustness and reliability of the quantum state transfer protocol and probe its potential limitations 
caused by various practical imperfections. We have therefore experimentally tested the above example of quantum state transfer using linear optics, see Fig.~\ref{fig:schema}. 
The source and target qubits were represented by polarization states of single photons generated in the process of spontaneous parametric downconversion. 
The probabilistic interaction $\hat{V}$ was realized by interference of the two photons on a partially polarizing beam splitter PPBS, 
which yields \cite{Starek16} $t_1=t_V$ and $t_{11}=2t_V^2-1$, where $t_V^2=0.334$ denotes the transmittance of the PPBS for vertically polarized photons. 
A tunable polarization quantum filter $\hat{G}$  was implemented by a combination of calcite beam displacers and wave plates, 
and the electrooptical feed-forward was implemented using a fiber interferometer and a fast phase modulator driven by a signal from a single-photon detector. For details, see Methods.

The state transfer was tested for $17$ different initial states of the target photon,
$|g\rangle_T=\cos\omega |0\rangle+\sin\omega|1\rangle$, with $\omega \in \{5, 10, 15, ..., 85\}$~deg. For each choice of $|g\rangle_T$ we have performed a full quantum process
tomography of the resulting single-qubit quantum channel $\mathcal{L}$ describing the state transfer from source qubit to the target qubit.
This quantum channel can be conveniently represented by a matrix $\chi$, which is a positive semidefinite operator on Hilbert space of two qubits.
Physically, the qauntum process matrix of a quantum channel $\mathcal{L}$ can be obtained by taking a pure maximally entangled Bell state
$|\Phi^{+}\rangle= \frac{1}{\sqrt{2}}(|00\rangle+|11\rangle)$ and sending one of the qubits through the channel $\mathcal{L}$.
A perfect state transfer corresponds to the identity channel whose matrix $\chi$ is equal to projector onto Bell state $|\Phi^{+}\rangle$.

\begin{figure}[!t]
	\centerline{\includegraphics[width=0.9\textwidth]{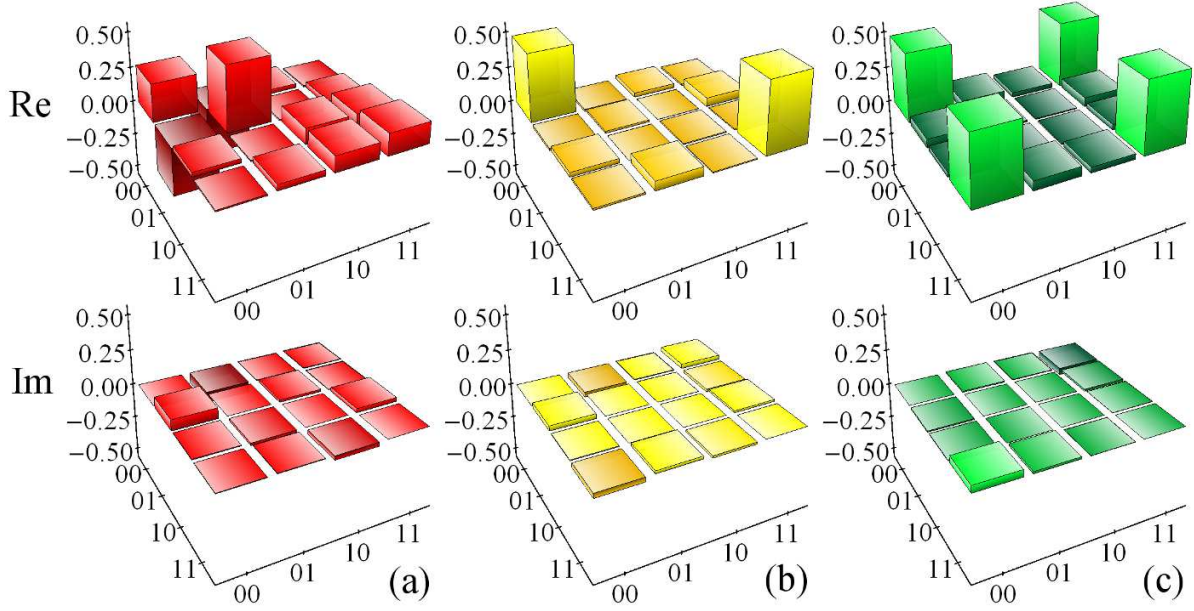}}
	\vspace*{3mm}
	\caption{The reconstructed channel matrices $\chi$  are plotted for $\omega = 55^\circ$ and for three scenarios:
		(a) both filtering and feed-forward switched off, (b) filter set on but feed-forward switched off, and (c)
		full implementation with both quantum filter and feed-forward on. The first and the second row show the real and imaginary parts 
		of the reconstructed matrix, respectively. For ease of comparison, all matrices are normalized such that $\mathrm{Tr}(\chi)=1$.}
	\label{fig:res1cut}
\end{figure}

In figure~\ref{fig:res1cut} we plot the reconstructed quantum channel matrices for $\omega=55^\circ$.
To show the importance of state filtration and feed-forward in our protocol, we first switch off
both of these operations, while accepting all coincidences of the detectors. This emulates the situation when we have information that the interaction $\hat{V}$
between the source and target qubit took place, but we do not perform any measurement on the source qubit (which may be e.g. inaccessible),
 and do not apply any operation to the target qubit. The resulting noisy quantum channel is shown in Fig. \ref{fig:res1cut}(a).
 If we switch on the fixed quantum filter $\hat{G}_{+}$ but keep the feed-forward switched off,
we obtain the quantum channel plotted in Fig. \ref{fig:res1cut}(b). Theory predicts that the fixed filtering should yield a dephasing channel represented by a diagonal operator
$\chi_{DC}=|00\rangle\langle00|+|11\rangle\langle 11|$, and our data are in very good agreement with this theoretical expectation.
Note that the dephasing channel is the best we can get without having access to results of measurements on the source qubit,
because the correlations present in the entangled state of  source and target qubits
destroy any phase coherence in the reduced density matrix of the target qubit. Finally, if we switch on also the feed-forward,
we achieve faithful state transfer, with the resulting channel being close to the identity channel, see Fig. \ref{fig:res1cut}(c).
In particular, compared to Fig. \ref{fig:res1cut}(b), the off-diagonal elements of the
channel matrix are recovered, as the feed-forward ensures preservation of quantum coherence between the computational basis states.

\begin{figure} [!t]
	\centerline{\includegraphics[width=0.55\textwidth]{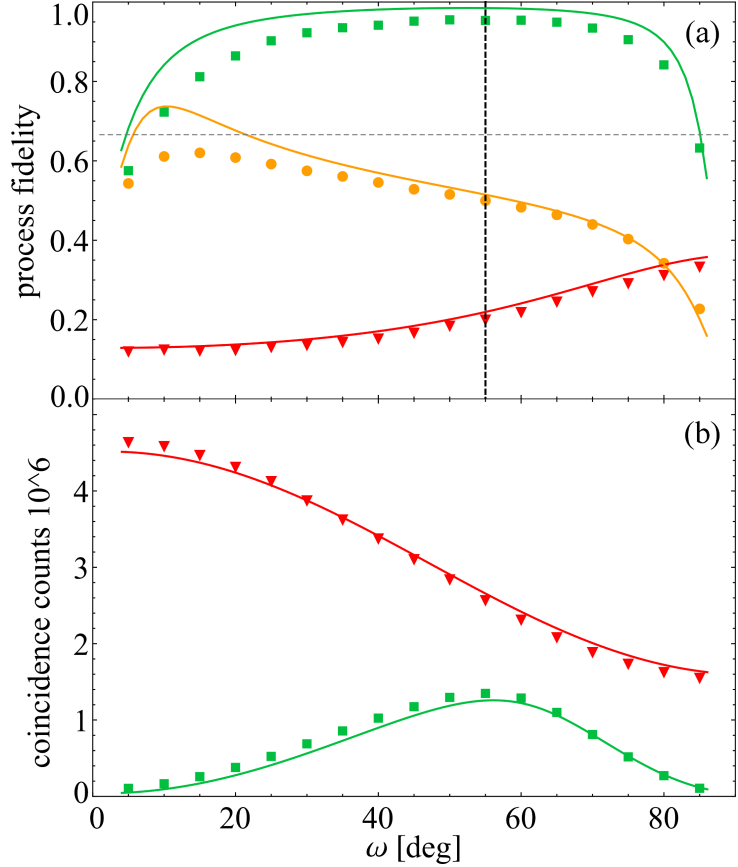}}
	\vspace*{3mm}
	\caption{Dependence of  quantum channel fidelity $\mathcal{F}$ (a) and two-photon coincidence counts (b)
	on initial polarization state of target qubit specified by angle $\omega$.
		The results are shown for three versions of the protocol:  full implementation with quantum filter and feed-forward ({\color{MYgreen} {\scriptsize $\blacksquare$}}),
		quantum filter set on but feed-forward switched off ({\color{MYyellow} {\large  $\bullet$}}), and both filtering and feed-forward switched off ({\color{red} {\normalsize $\blacktriangledown$}}).
		The horizontal dashed line shows the classical measure-and-prepare bound $\mathcal{F}=2/3$.
		The vertical dashed line indicates the setting $\omega=55^\circ$ for which the channel matrices are plotted in Fig. \ref{fig:res1cut}. 
		The solid lines indicate predictions of a theoretical model that accounts for imperfections of the partially polarizing beam splitter where the source and target photons interfere.}
	\label{fig:res1}
\end{figure}

A quantitative characterization of performance of the quantum state transfer protocol in dependence on the angle $\omega$ specifying the initial polarization state of the target photon
is provided in Fig.~\ref{fig:res1}. Figure \ref{fig:res1}(a) shows the channel fidelity $\mathcal{F}$ 
defined as a normalized overlap of the channel matrix $\chi$ and the Bell state $|\Phi^{+}\rangle$.
We achieve a high fidelity with maximum $\mathcal{F} = 95.8~\%$ at $\omega=50^\circ$. The experimental imperfections which reduce the fidelity below $1$
include deviations of the interaction from ideal one (\ref{VPPBS}) and decoherence of the source and target qubits. 
The imperfections depend on physical realization of the interaction, but they sufficiently emulate real problems 
which appear for the transfer of quantum states. Heavy quantum filtering can be required in extreme cases, which makes the transfer more sensitive to even small deviations and decoherence.     

Figure \ref{fig:res1}(b) simultaneously illustrates the dependence of the success rate of the protocol on the initial state of the target qubit. For each $\omega$
we plot the sum of all measured two-photon coincidences, which is proportional to the success probability $p$. Since the same measurements were carried out for each $\omega$
and the measurement time was kept constant, the data for various $\omega$ are directly comparable. For reference, we plot also the total coincidences recorded without filtering.
The success rate of the protocol is maximized at $\omega=55^\circ$. We have performed numerical optimization of the success probability (\ref{pdefinition})
for the ideal protocol and we have found that $p$ is maximized at $\omega=55.2^\circ$, which is in excellent agreement with our experimental observations.

\subsection*{Optimality of the protocol}

\begin{figure}[!t!]
	\centerline{\includegraphics[width=0.6\textwidth]{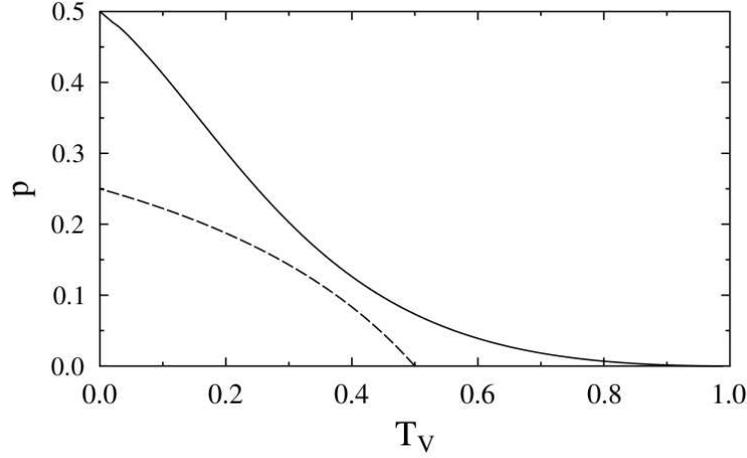}}
	\caption{Success probability $p$ of the optimal protocol (solid line) and success probability $\tilde{p}$ of the simplified protocol avoiding filtering (dashed line)
		are plotted in dependence on the transmittance $T_V=t_V^2$ of the partially polarizing beam splitter PPBS.}
	\label{fig:psucc}
\end{figure}

To maximize the success probability of the protocol we can, in addition to the initial state of the target qubit $|g\rangle_T$, 
 optimize also the basis $|\pi\rangle_S$, $|\pi^\perp\rangle_S$ in which the source qubit is measured. 
Since the operator  $\hat{V}$ given by Eq. (\ref{VPPBS}) 
is diagonal in the computational basis, phase shifts do not play any role in the optimization. Therefore we can, without loss of any generality,
restrict our analysis to real qubit states $|g\rangle_T=\cos\omega|0\rangle_T+\sin\omega|1\rangle_T$ 
and $|\pi\rangle_S=\cos\kappa |0\rangle_S+\sin\kappa|1\rangle_S$, which represent linearly polarized states of single photons.
Here $\kappa$ is an angle that parametrizes the measurement on the source qubit and $|\pi^\perp\rangle_S=\sin\kappa |0\rangle_S-\cos\kappa|1\rangle_S$. 
We focus on the interferometric coupling on a PPBS specified by $t_1=t_V$ and $t_{11}=2t_V^2-1$. 
We first fix $\omega$ and $t_V$ and maximize the success probability $p$ over $\kappa$. 
Our numerical calculations indicate that for any $t_V$ and $\omega$ it is actually optimal to set $\kappa=\pi/4$ and measure the source qubit in the balanced superposition 
basis $|\pm\rangle_S$, as implemented in our experiment. For a given fixed interaction strength $t_V$ we further perform optimization of $p$ over $\omega$ and
the success probability of the resulting optimal protocol is plotted in Fig.~\ref{fig:psucc}.
For comparison, the dashed line in Fig.~\ref{fig:psucc} shows the success probability $\tilde{p}$ 
of a simplified protocol with $|g\rangle_T$ and $|\pi\rangle_S$ chosen such that the filtering is not required and $\hat{G}$ is proportional to a unitary operation (see Methods). 
Such simplified protocol can be constructed only if the interaction is sufficiently strong, $t_V^2<1/2$, and its success probability $\tilde{p}$ is strictly smaller than $p$.

\section*{Discussion}

In summary, we have proposed and experimentally verified a conditional transfer of single qubit state through weak trace-decreasing interaction using optimal measurement 
and irreducible quantum filtering in the feed-forward loop. The method is feasible, robust and universally applicable. It can be suitable for broad class 
of qubit transfers in hybrid information processing, including atomic, solid state and optical qubits.

\section*{Methods}

\textbf{Quantum filter.} Any quantum filter $\hat{G}$ can be decomposed into a sequence $\hat{G}=\hat{U}_1\hat{D} \hat{U}_2$, where $\hat{U}_{1,2}$ are unitary operations and the operator $\hat{D}$ is diagonal in the computational basis,
$\hat{D}=|0\rangle\langle 0|+\lambda |1\rangle \langle 1|$, where  $\lambda\in (0,1]$ is an attenuation factor. Such diagonal filter can be implemented by a
weak measurement in the computational basis. This can be accomplished by coupling the target qubit to a meter qubit M prepared initially in state $|0\rangle_M$.
The required coupling reads
\begin{eqnarray}
|0\rangle_T|0\rangle_M &\rightarrow & |0\rangle_T|0\rangle_M, \nonumber \\
|1\rangle_T|0\rangle_M &\rightarrow & \lambda|1\rangle_T|0\rangle_M+\sqrt{1-\lambda^2}|1\rangle_T|1\rangle_M.
\end{eqnarray}
Successful implementation of the filter $\hat{D}$ is heralded by projection of meter qubit onto state $|0\rangle_M$.
\\

{\bf Real-time feed-forward control.} We can improve the success probability by exploiting both outcomes $|\pi\rangle$ and $|\pi^\perp\rangle$ of projective measurement on the source qubit. Since the filter $\hat{G}$
depends on the measurement outcome, this requires a real-time feed-forward which ensures that a correct quantum filter $\hat{G}$ or $\hat{G}_\perp$ is aplied to the target qubit,
where $\hat{G}_\perp$ denotes a quantum filter corresponding to projection of source qubit onto $|\pi^\perp\rangle$. The total success probability of state transfer
is then given by a sum of two contributions,
\begin{equation}
p=\frac{1}{|N|^2}+\frac{1}{|N_\perp|^2},
\label{pdefinition}
\end{equation}
where $N_\perp$ is the normalization factor in $\hat{G}_\perp$. The probabilistic nature of the quantum state transfer protocol is the price to pay
for a faithful transfer of all states with unit fidelity. The quantum filter $\hat{G}$ is an essential part of the protocol and the average fidelity of states (\ref{phiT}) obtained without filtering
may even drop below the limit of $2/3$, which is achievable by a purely classical measure-and-prepare strategy.
\\

\textbf{Linear-optical emulation of interaction.} We demonstrate the state transfer protocol using linear optics and qubits encoded into polarization states of single photons,
where the computational basis states $|0\rangle$ and $|1\rangle$ correspond to horizontally and vertically polarized photon, respectively.
We use this platform as a suitable testbed for a proof-of-principle verification of our protocol, whose applicability is universal and by no means limited to photonic qubits.
A suitable non-trivial interaction between single photons is provided by two-photon interference on a polarizing beam splitter PBS that fully transmits horizontally polarized photons
and fully reflects vertically polarized photons. If we post-select on presence of a single photon
in each output port of PBS, we obtain the so-called quantum parity check \cite{Pittman02,Hofmann02,Okamoto09},
\begin{equation}
\hat{V}_{QPC}= |00\rangle\langle 00|- |11\rangle \langle 11|.
\end{equation}
If the target qubit is initially prepared in a superposition state $|+\rangle=\frac{1}{\sqrt{2}}(|0\rangle+|1\rangle)$ and the source qubit is measured in the superposition basis 
$|\pm\rangle=\frac{1}{\sqrt{2}}(|0\rangle\pm|1\rangle)$, then we find that the state of target qubit before filtering reads $\frac{1}{2}(\alpha|0\rangle \mp \beta |1\rangle)_T$,
hence the state transfer can be accomplished by a unitary feed-forward operation which applies
a $\pi$ phase shift to the state $|1\rangle_T$ iff the source qubit is projected onto state $|+\rangle$ \cite{Mikova13}. To make our study more generic 
and test our procedure in the regime of weakly coupled qubits, we utilize interference on a partially polarizing beam splitter (PPBS) that is still fully transmitting for
horizontally polarized photons and  only partially reflecting for  vertically polarized photons, with corresponding amplitude transmittance $t_V$. 
Conditional on presence of a single photon at each output port of PPBS \cite{Starek16}, 
the interference results in the two-qubit transformation (\ref{VPPBS}) with $t_1=t_V$ and $t_{11}= 2t_V^2-1$. 
\\

\textbf{Experimental set-up.}
Time correlated photon pairs are generated in the process of frequency degenerate parametric down-conversion and fed to the input of the experimental setup shown in Fig.~\ref{fig:schema}.
Arbitrary input states of source and target qubit can be prepared by a combination of quarter-wave plate (QWP) and half-wave plate (HWP).
The qubits interact at a partially polarizing beam splitter (PPBS) with transmittance $T_V=t_V^2=0.334$.
Subsequently, the source qubit is measured in the basis of diagonal linear polarizations
using a single-photon polarization detection block consisting of a half-wave plate, polarizing beam splitter and two single photon detectors.
The polarization filter $\hat{G}_{+}=\hat{U}_2\hat{D}\hat{U}_1$ on the target photon was implemented with the help of
calcite beam displacers (BD) and half-wave plates, see the inset in the upper part of Fig.~\ref{fig:schema}.
The half-wave plates at the input and output of the filter implement the unitary operations $\hat{U}_{1}$ and $\hat{U}_{2}$.
Selective attenuation of vertical or horizontal polarization
is implemented by a suitable rotation of half-wave plates inserted inside an interferometer formed by the two calcite beam displacers.
Since a beam displacer introduces a transversal spatial offset between vertically and horizontally polarized beams, these two polarization components
become spatially separated inside the interferometer and can be individually addressed \cite{Jeffrey04,Peters05,Micuda14}.

The conditional $\pi$ phase shift on the target qubit is applied by means of an active electro-optical feed-forward loop \cite{Giacomini02,Prevedel07,Mikova12,Mikova13}.
To facilitate its experimental realization,
we couple the target photon into fiber-based Mach-Zehnder interferometer (MZI), thus converting the polarization qubit into qubit encoded in which way information. 
A sufficiently long optical fiber delays the photon and provides time necessary for processing the electronic signal produced by the single photon detector D3
whose click indicates projection of source qubit onto state $|-\rangle$.  This signal is amplified and fed to an integrated lithium-niobate phase modulator PM1
inserted in one arm of the fiber interferometer, thereby applying the conditional $\pi$ phase shift to state $|1\rangle_{T}$.
The required stability of the setup was reached by thorough isolation from the environment and, simultaneously,
by active phase stabilization of MZI every 1.5~s to reduce phase drifts caused by remaining temperature and air-pressure fluctuations below $2^\circ$ during the measurement.

The experimental imperfections which reduce the fidelity below $1$
include  partially distinguishable photons, imperfect retardation of wave plates, interference visibility lower than one, and imperfection of the partially polarizing beam splitter, 
which was not perfectly transmitting for horizontally polarized photons, and the experimentally determined horizontal transmittance was $T_H=0.983$.
This latter effect plays dominant role when the initial state of the target photon becomes close to horizontally
or vertically polarized one. Heavy quantum filtering is required in such cases, as illustrated in Fig.~\ref{fig:res1}(b), which makes the protocol more sensitive to
parasitic coupling in horizontal polarization.
\\

\textbf{Simplified protocol without filtering.}
We can attempt to choose $|g\rangle_T=\cos\omega|0\rangle_T+\sin\omega|1\rangle_T$ and $|\pi\rangle_S=\cos\kappa |0\rangle_S+\sin\kappa|1\rangle_S$ so as to maximally simplify the protocol. 
In particular, we can try to completely avoid the need for quantum filtering. It follows from our theoretical analysis of the protocol that quantum filtering is not required provided 
that the states $|\phi_0\rangle$ and $|\phi_1\rangle$  are orthogonal and have the same norm,
\begin{equation}
\langle \phi_1|\phi_0\rangle=0, \qquad \langle \phi_0|\phi_0\rangle=\langle \phi_1|\phi_1\rangle.
\label{orthoconditions}
\end{equation}
For the two-qubit interaction (\ref{VPPBS}) with $t_1=t_V$ and $t_{11}=2t_{V}^2-1$
the orthogonality condition can be satisfied only if $t_V^2<1/2$, i.e. only if the interferometric coupling of the two photons is sufficiently strong.
The parameters $\omega$ and $\kappa$  are uniquely determined by the conditions (\ref{orthoconditions}) and we obtain
\begin{equation}
\tan\omega=\tan\kappa=\frac{1}{\sqrt{1-2t_V^2}}.
\end{equation}
Success probability of this simplified protocol then reads
\begin{equation}
\tilde{p}=\frac{1-2t_V^2}{4(1-t_V^2)}.
\end{equation}


\section*{Acknowledgements}
This work was supported by the Czech Science Foundation (GA13-20319S). V.K. acknowledges support by Palacky University (IGA-PrF-2015-005 and IGA-PrF-2016-009).
We thank M. Dudka for discussions about feed-forward loop electronics.

\section*{Author contributions statement}
R. Filip and J. Fiur\'a\v sek developed the theory and theoretical schemes. M. Mikov\'{a} and M. Je\v{z}ek designed the experimental setup, 
M. Mikov\'{a} performed the experiment with contributions from V. Kr\v{c}marsk\'{y}, I. Straka, and M. Mi\v{c}uda, and M. Je\v{z}ek 
and M. Du\v{s}ek supervised and coordinated the experiment. All authors discussed the experimental results. 
R. Filip and J. Fiur\' a\v sek wrote the manuscript with input from all authors.

\section*{Additional information}
The authors declare no competing financial interests. Correspondence and requests for material should be adddressed to R.F. filip@optics.upol.cz.

\end{document}